\documentclass[twocolumn,aps,pra,floats,superscriptaddress,showpacs]{revtex4}
\usepackage{amsmath}
\usepackage{epsfig}
\usepackage{graphicx}
\usepackage{dcolumn}
\usepackage{bm}
\usepackage{color}

\def\gapp{\lower.35em\hbox{$\stackrel{\textstyle>}{\sim}$}}
\def\lapp{\lower.35em\hbox{$\stackrel{\textstyle<}{\sim}$}}

\begin{document}

\title{{Electrical and thermal control of Fabry-P\'{e}rot cavities mediated by Casimir
forces}}
\author{Lixin Ge}
\email{lixinge@hotmail.com}
\affiliation{School of Physics and Electronic Engineering, Xinyang Normal University,
Xinyang 464000, China}
\author{Bingzhong Li}
\affiliation{School of Physics and Electronic Engineering, Xinyang Normal University,
Xinyang 464000, China}
\author{Hao Luo}
\affiliation{School of Physics and Electronic Engineering, Xinyang Normal University,
Xinyang 464000, China}
\author{Ke Gong}
\affiliation{School of Physics and Electronic Engineering, Xinyang Normal University,
Xinyang 464000, China}
\date{\today }

\begin{abstract}
Dynamic tuning of optical cavities is highly desired in many photonic systems. Here, we
show that Fabry-P\'{e}rot(FP) cavities can be actively controlled by the Casimir force.
The optical FP cavities consist of a gold nanoplate confronted to an electrical-connecting multi-layer substrate in a liquid environment. The gold nanoplate can be stably suspended due to the balance of repulsive and attractive Casimir forces. Moreover, the suspension distance are modulated strongly by the electric gating or temperature of the system. As a result, we could shift the resonant wavelengthes of the cavities with tens of nanometers at optical frequencies. Finally, we analyze the influence of
Brownian motion on the equilibrium distances. Due to the high $Q$-factor
of the FP cavities, our proposed system offers a remarkable
platform to experimentally investigate the thermal Casimir effect at
sub-micrometer separations
\end{abstract}

\maketitle



\section{Introduction}

The Casimir force between two perfect metallic plates, predicted by Hendrik
Casimir in 1948, is a macroscopic quantum effect resulting from the
zero-point fluctuations in vacuum \cite{Cas:48}. Latter, this quantum effect
was generalized by E. M. Lifshitz to include frequency-dependent dielectrics
and finite temperatures \cite{Lif:55}. The Casimir forces between two
objects consisted by the same materials are generally attractive. In the
past two decades, great effort has been devoted to the search for Casimir
repulsions in the vacuum environment\cite{Woo:16, Ken:02, Zha:09, Lev:10, Rod:14, Nie:13, Li:21, Cam:22}, yet lack of experimental verifications due to the strict constrains.
By contrast, the Casimir repulsions have been experimentally achieved between two liquid-separated
objects (labelled 1 and 2) when the permittivity satisfies $%
\varepsilon_1(i\xi) >\varepsilon_{\mathrm{liq}}(i\xi) > \varepsilon_2(i\xi)$
for a vast range of frequency \cite{Mun:09}, where $\varepsilon_{%
\mathrm{liq}}(i\xi) $ is the permittivity of the intervening liquid
evaluated with imaginary frequency $\omega =i\xi $. Interestingly, stable
suspensions mediated by Casimir repulsions were reported in different configurations\cite{Zha:19, Dou:14, Est:15, Liu:16, Ye:18, Est:22}.

Recently, an new concept for tunable FP cavities has been proposed by Esteso
et al.\cite{Est:19}, based on the Casimir
force. The FP cavities play a crucial role in optical
spectroscopy and find extensive applications\cite{Vau:17}. For instance, the FP cavities consisted of metal-insulator-metal have been received considerable interest in nanophotonics, due to their excellent performances on strong light-matter interactions \cite{Cal:20, Liu:10,Den:15}. In
general, the resonances of FP cavities are fixed once
the samples are fabricated \cite{Est:19}. The dynamic tuning of optical FP cavities through the Casimir force, particularly using external stimuli such as electric gating, and temperature, remains largely unknown in this field.

Tunable Casimir forces can be realized by changing the dielectric response of the materials through external stimuli, e.g., electric gating \cite%
{Ge:20b, Gon:22}, magnetic fields \cite{Jia:19, Zen:11, Wan:06}, optical lasers \cite{Che:07, Cha:11}, etc. Another scheme to dynamically tune the Casimir forces is based on the change of temperature \cite{Yam:08, Gal:09,
Bos:18, Ge:22}. Generally, thermal effect on the Casimir forces is
weak \cite{Mat:00, Sus:11}. For a vacuum gap, the
thermal Casimir effect is observable only when the separation is large (e.g., over
three micrometers)\cite{Sus:11}. Such large separation severely affects its applications.
Recently, a strong thermal Casimir effect based on graphene
sheets was revealed at sub-micrometer scales \cite{Liu:21, Abb:17, Bim:17,
Khu:18}. The temperature dependence of Casimir forces for graphene is
attributed to two different mechanisms. The first one is the thermal
fluctuation, as illustrated by the implicit term in \cite{Kli:15}. The second one relies on the fact that the
dielectric response of graphene is temperature dependence (the explicit term
in \cite{Kli:15}). This kind of temperature modulation can be manifest at
shorter separations.

In this study, we aim to dynamically tune FP cavities by manipulating the Casimir forces. Our system comprises a gold nanoplate and a Teflon-coated metal-oxide-semiconductor (MOS) substrate. Notably, the Casimir forces acting on the suspended gold nanoplate exhibit a strong dependence on both the gating voltage and the temperature. As a result, the equilibrium separation of the gold nanoplate undergoes significant alterations. The resonant wavelength of the optical FP cavities can be shifted for tens of nanometers. These remarkable shifts can be accurately detected using state-of-the-art experimental techniques. At the end, the Brownian motion of nanoplates is taken into account, and our study presents an accurate approach for measuring thermal Casimir forces at sub-micrometer separations via spectroscopic of the cavities.

\section{Theoretical models}

\begin{figure}[tbp]
\centerline{\includegraphics[width=8.0cm]{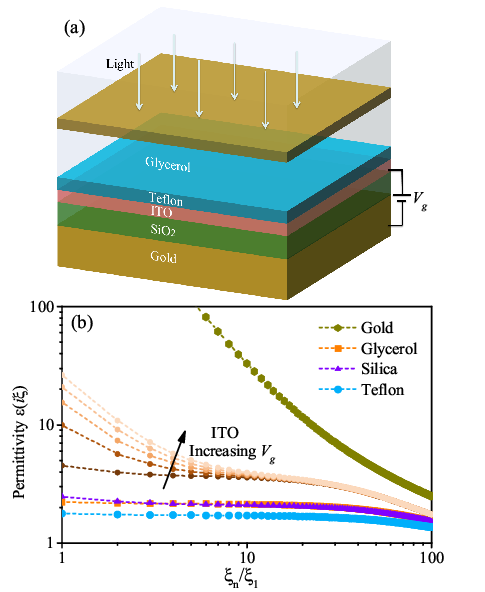}}
\caption{(color online) (a)Schematic view of the system. The incident light is a plane wave, and the two parallel
reflecting mirrors are consisted of the gold nanoplate and the gold
substrate. (b) The dielectric response of materials evaluated in imaginary
frequency. The plot depicts the permittivity of the accumulation layer of ITO, where the gating voltage increases from 0 to $V_b$ with a step size of $V_b/4$.
}
\label{Fig:fig1}
\end{figure}

Figure 1(a) illustrates the schematic of the system under study. In
this setup, a gold nanoplate is suspended in a liquid of glycerol. The materials of Teflon, indium tin oxide(ITO), silica, and
glycerol, are almost transparency at optical frequencies. Hence, the two parallel reflecting
mirrors of the optical FP cavities are the suspended gold nanoplate and the
gold substrate. There exists an electrical connection between the gold
substrate and the ITO layer, which can be controlled by a gating
voltage.

Considering the in-plane dimension of nanoplate is much larger than the
separation, a proximity force approximation is applied for the
calculations. The Casimir force is calculated by $F_{c}=-\partial
E_{c}(d)/\partial d$, where $E_{c}(d)$ is the Casimir energy between the
nanoplate and the multilayer substrate:\cite{Zha:19}
\begin{equation}
\frac{E_{c}(d)}{A}=k_{b}T \overset{\infty }{\underset{n=0}{\sum}}^{\prime} \int \frac{%
d^{2}\mathbf{k_{\Vert }}}{(2\pi )^{2}}\log \det \left[ 1-\mathbf{R_{1}}\cdot
\mathbf{R_{2}}e^{-2 K_{n}d}\right] ,
\end{equation}%
where $A$ represents the in-plane area, $k_{B}$ is the Boltzmann's constant, $T$ is the temperature,$d$ is the separation, the prime denotes a pre-factor 1/2 for $n$=0,
 $\mathbf{k_{\parallel }}$ is the parallel wavevector, $K_{n}=\sqrt{k_{\parallel }^{2}+\varepsilon _{liq}(i\xi_n
)\xi _{n} ^{2}/c^{2}}$ is the vertical wavevector in the liquid, $\xi
_{n}=2\pi \frac{k_{b}T}{\hbar }n (n=0,1,2,3\ldots )$ is the discrete
Matsubara frequencies, $\hbar$ is the reduced Planck's constant,
$c$ is the speed of light in vacuum. $\mathbf{R}_{1,2}$ is the $2\times 2$
reflection matrix, given by
\begin{equation}
\mathbf{R_{j}}=\left(
\begin{array}{cc}
r_{j}^{s} & 0 \\
0 & r_{j}^{p}%
\end{array}%
\right) ,
\end{equation}%
where $r_{j}^\alpha$ with $j$=1 and $j$=2 are the reflection coefficients for the
upper and lower layered structures, and the superscripts $\alpha=s$ and $p$
correspond to the polarization of transverse electric ($\mathbf{TE}$) and
transverse magnetic ($\mathbf{TM}$) modes, respectively. The reflection
coefficients are associated with the layer
thicknesses and permittivity of materials, which can be
calculated by a transfer matrix method (TMM)\cite{Ge:20a}.

The generalized Drude-Lorentz model is applied for the gold\cite{Ge:20a}. The dielectric models and parameters for the materials of Teflon, silica, and glycerol are adopted from the recent literatures \cite%
{Moa:21,Seh:17}.  The permittivity for the ITO film is evaluated by the
Kramers-Kronig relationship:
\begin{equation}
 \varepsilon (i\xi_n )=1+\frac{2}{\pi }%
\int_{0}^{\infty }\frac{\mathrm{Im}[\varepsilon (x)]x}{x^{2}+\xi_n ^{2}}dx.
\end{equation}
The absorption of the ITO materials (imaginary part)is constructive by the sum of the Drude model \cite{Kra:12} and the Tauc-Lorentz model \cite{Jel:96,Ban:12}.
Under electrical-bias, the ITO layer is divided into two distinct regions: the background layer and
the accumulation layer. The carrier densities in these two layers are represented as $N_b$ and $%
N_a$, respectively. The magnitude of $N_b$ is fixed upon fabrication,
whereas the magnitude of $N_a$ can be adjusted by varying the gating
voltage. As given in previous literatures, the accumulation layer possesses
a homogeneous carrier density, and its thickness is determined by\cite{Gon:22, Yi:13}:
\begin{equation}
L_{a}=\frac{\pi }{\sqrt{2}}\sqrt{\frac{k_{B}T\varepsilon _{0}\varepsilon _{%
\mathrm{ITO}}}{N_{b}e^{2}}},
\end{equation}
where $e$ is the electron charge, $\varepsilon _{0}$ is the vacuum permittivity, $%
\varepsilon _{\mathrm{ITO}}$=9.3 is the static permittivity of ITO. In this
work, The thickness of the ITO layer is fixed to be 5 nm, and the background
carrier density $N_\mathrm{b}$= $10^{19}$ cm$^{-3}$. We have $L_\mathrm{a}$%
=2.56 nm at $T$=300 K. On the other hand, the analytical expression of $N_{%
\mathrm{a}}$ is given by \cite{Gon:22}:%
\begin{equation}
N_{a}=N_{b}+\frac{\varepsilon _{0}\varepsilon _{s}V_g}{eL_{s}L_{a}},
\end{equation}
where $\varepsilon _{\mathrm{s}}$=3.9 denotes the static dielectric constant
of silica, $V_g$ is the applied voltage, and it should be
smaller than the breaking down voltage $V_b$=$E_bL_s$, where we assume the breakdown field $E_b$=30 MV/cm \cite{Pap:15, Sir:07}.

The permittivity evaluated in the imaginary frequency is presented in Fig. 1(b). Notably, the dielectric functions of the ITO in the accumulation layer exhibit significant variations with increasing voltages in the infrared frequency range. Teflon possesses the lowest permittivity, making it particularly desirable for Casimir repulsions at small separations. Although the permittivity of glycerol is close to that of silica at infrared and visible frequencies, its static permittivity, about 42.4, at zero frequency is much larger than that of silica.

\begin{figure}[tbp]
\centerline{\includegraphics[width=8.0cm]{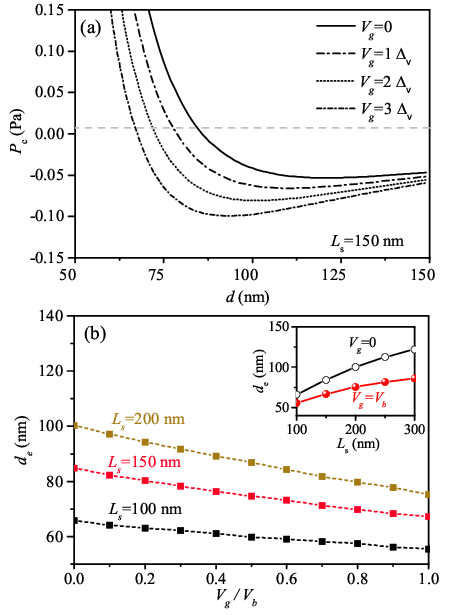}}
\caption{(color online) (a) The Casimir pressures under different gating
voltages, where $\Delta_\mathrm{v}=V_b/3$, where the breakdown voltage is 450 V. The positive (negative) sign of the
pressure corresponds to the repulsive (attractive) force. The gray dashed line represents the pressure generated from the gravity and buoyancy. (b)The equilibrium
distances as a function of applied voltages. The inset show the equilibrium
distances as a function of the thickness of silica layer.}
\label{Fig:fig2}
\end{figure}
\begin{figure}[htbp]
\centerline{\includegraphics[width=8.0cm]{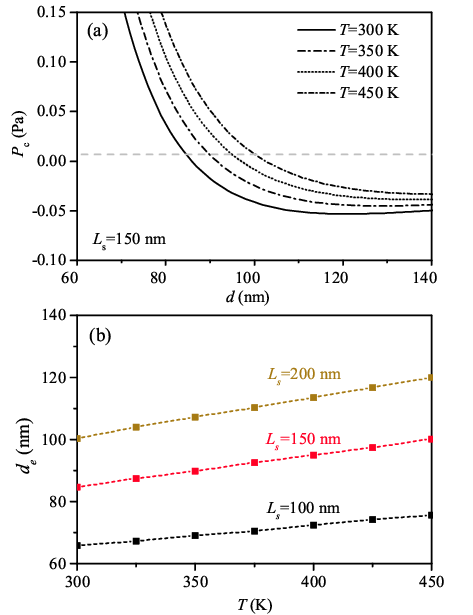}}
\caption{(color online) (a) The Casimir pressures for different
temperatures, where there is no applied voltage. The gray dashed line represents the pressure generated
from the gravity and buoyancy. (b)The equilibrium distances as a function
of temperature. }
\label{Fig:fig3}
\end{figure}

\section{Results and discussions}
The Casimir pressure of suspended gold nanoplate versus the separation is shown in Fig. 2(a). The thickness of the gold nanoplate, Teflon, ITO, and silica layers are set to be 40, 10, 5, and 150 nm, respectively. The results demonstrate a significant modulation of the Casimir pressure by the gating voltage. At small separations, the Casimir pressure is repulsive, while it turns to be an attractive force for larger separations. A separation for zero pressure, known as the Casimir equilibrium, is identified at a specific separation.  With increasing the voltages, the pressure tends to be
attractive, and the separation for the Casimir equilibrium decreases correspondingly. In the absence of voltage, the Casimir equilibrium appears at approximately 85 nm, which decreases by about 18 nm as the voltage approaches to $V_b$. For a gold nanoplate with thickness of 40 nm, the pressure resulting from gravity and buoyancy is estimated to be 0.007 Pa \cite{Ge:20a}, and is represented by the dashed grey lines in Fig. 2(a). The equilibrium distance, denoted as $d_e$, is established through the delicate balance among the Casimir force, gravity, and buoyancy. The $d_e$ is slightly smaller than the separation of Casimir equilibrium.

\begin{figure*}[htbp]
\centerline{\includegraphics[width=15.0cm]{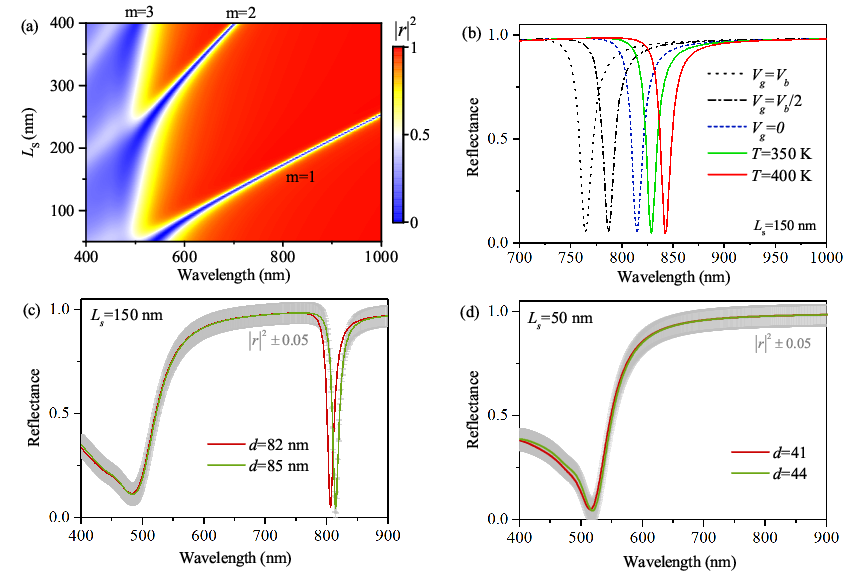}}
\caption{(color online) (a)The reflectance of FP cavity
varies with the layer thickness of silica, where the separation $d$ is fixed at 60 nm. (b)The modulation of reflectance through the applied voltage and temperatures. (c) and (d) show the reflection spectra in and out of the equilibria. The equilibrium distances for $L_s$=150 nm and $L_s$=50 nm are 85 nm and 44 nm, respectively. The
assumed error of the reflectance with $\pm$ 0.05 is given for the equilibrium separation represented by the grey bars. }
\label{Fig:fig4}
\end{figure*}

The equilibrium distance as a function of applied voltage is depicted in Fig. 2(b). The findings demonstrate a decline in the equilibrium distance as the voltage increases, ranging from 0 up to $V_b$. The equilibrium distance is also influenced by the thickness of the silica layer, as shown for $L_s$=100, 150 and 200 nm. The inset of Fig. 2(b) reveals that increasing the layer thickness, would results in elevated equilibrium distances, and the difference of $d_e$ between $V_g=0$ and $V_g= V_b$ expands. At $L_s$ = 300 nm, the difference of $d_e$ for the $V_g=0$ and $V_g= V_b$ reaches nearly 36 nm. The calculations suggest that the equilibrium distance, so as the resonant length, of the FP cavities can be effectively modulated by the tunable Casimir forces through electrical gating.

The Casimir pressure is also strongly dependent on the temperature, as shown in Fig. 3(a). The layer thicknesses of the materials are kept the same as those in Fig. 2(a). To manifest the temperature effect on the Casimir force, the applied voltage is assumed to be zero. We find that the Casimir pressure
tends to be more repulsive as the temperature increases,
and the variation of $d_e$ is near 10 nm when the temperature increases from 300 to 400 K. The Casimir pressure as a function of temperature is shown in Fig. 3(b) under different $L_s$. Again, the value of $d_e$ increases when the layer thickness $L_s$ increases from 100 to 200 nm. The variation of $d_e$ with respect to $T$ is almost linear as reported in \cite{Est:16}. Such effective modulation of $d_e$ due to the temperature has been proposed for detection of thermal Casimir effect \cite{Rod:10}. However, the optical resonances have not been employed in the literature\cite{Est:16, Rod:10}.

The thickness of silica needs to be carefully designed.
The contour plot of the reflectance via silica thickness and wavelength $\lambda$ is shown in Fig. 4(a), wherein the separation $d$ is fixed at
60 nm. The reflectance is calculated by the TMM in the real frequency \cite%
{Zha:13}. The results reveal that different resonant modes are excited when
the thickness $L_s$ varies from 50 to 400 nm. A fundamental mode
with $m$=1 is excited in the visible regime when the layer thickness $L_s$
is in proper range. As the thickness increases to a higher value, e.g., 300 nm, other high-order modes are presented.

\begin{figure}[htbp]
\centerline{\includegraphics[width=8.0cm]{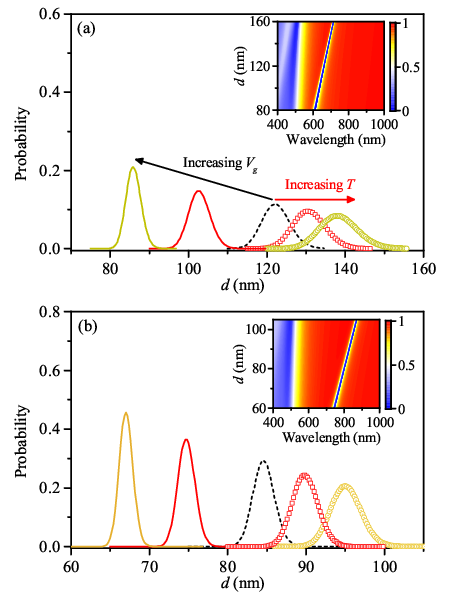}}
\caption{(color online) The probability of suspension positions due to the
Brownian effect with (a) $L_\mathrm{s}$=300 nm and (b) $L_%
\mathrm{s}$=150 nm. The increasing voltages are 0, V$_\mathrm{b}$/2, and V$_%
\mathrm{b}$. The increasing temperatures are 300, 350 and 400 K. The dashed
lines represent the initialing state with $V_g$=0 and $T$=300 K. The insets in (a) and (b) show the corresponding contour maps of the reflectance.
}
\label{Fig:fig5}
\end{figure}

Figure 4(b) show the reflection spectra under different external stimuli, where the nanoplate is suspended at the equilibrium
distance $d_e$. The resonance of the FP cavities is modulated efficiently by the applied voltage and temperature. The equilibrium distances are about 85, 75 and 67 nm for the applied voltage are 0, $V_b/2$ and $V_b$, respectively. On the other hand, the equilibrium distances increase to 90 and 95 nm for the temperatures 350 and 400 K, respectively. Hence, the resonant dip has a blue shift for increasing the voltages, while it has a red shift for increasing the temperature. The results show that the giant shifting of the resonances over tens of nanometers, is achieved by electrical gating or temperatures.

In a real configuration, the imperfection of the experiments
(e.g., misalignment, surface roughness, electrostatic forces...) may exist, and the errors of the reflectance with $\pm$ 0.05 (gray lines) should be introduced as indicated in \cite{Est:19}. Here, the high $Q$-factor of optical
cavities provides an avenue for accurate spectroscopic measurements. For instance, the resonant wavelength without the electrical gating appears at about 810 nm, where the equilibrium separation is 85 nm. When the separation is out of equilibrium with $d$=82 nm, the resonant wavelength can be clarify by the spectroscopy,  as shown in Fig. 4(c). The calculated results indicate that such small errors have limited influences as long as the spectra having high $Q$-factors. For thickness $L_s$=50 nm, the equilibrium separation is 44 nm, and the shift of the reflection spectrum could not clarify in Fig. 4(d). This is because the resonance of the cavities appear at the lossy regime ($\lambda$ is smaller than about 550 nm) with $L_s$=50 nm, the change of reflection spectrum due to the the variation of separation could not be detected by the spectroscopy at such configuration.

The Brownian motion should be considered when the suspended
nanoplate is finite size. The random Brownian motion happens at both lateral and
vertical directions. The lateral Brownian motion does not affect the optical
resonances of FP cavities. However, the vertical Brownian motion could make
the stable suspension of nanoplates out of equilibrium. The normalized probability of the suspension distance due to the
vertical Brownian effect is given by \cite{Var:11}:
\begin{equation}
\rho (d,T)=\frac{\text{Exp}{[-U(d)/k_{B}T]}}{\int\limits_{0}^{\infty }\text{%
Exp}{[-U(d)/k_{B}T]\mathrm{\partial }d}}
\end{equation}
where $U(d)=E_c(d)+F_{\mathrm{GB}}d$ is the total energy of the suspended
nanoplate, where $F_{\mathrm{GB}}$ is the sum of the gravity and buoyancy
forces \cite{Ge:20a, Var:11}. Here, we consider the gold nanoplate with area $A$=20 $\mu \mathrm{m}\times$ 20 $\mu
\mathrm{m}$ (see the experiment samples in \cite{Zha:19}). The normalized probability with respect to the separation is shown in Figs. 5(a) and
5(b), where we set $L_s$=300 and 150 nm, respectively. The higher of the
probability at the equilibrium distance, the narrower the suspension
distribution. The peak probability $\rho (d_e,T)$ for $L_s$=150
nm is almost 2 times larger than that of $L_s$=300 nm,
indicating a stronger stiffness at the quantum trapping. Overall, the probability $%
\rho (d_e,T)$ increases with increasing the applied voltage. By
contrast, the probability $\rho (d_e,T)$ decreases slightly with increasing
the temperature from 300 to 350 and 400 K. The distribution functions could be overlapping with each other at an intervening 50 K. The average of
separations would be \cite{Var:11}:
\begin{equation}
\bar{d}=\frac{\int\limits_{0}^{\infty }d\cdot \text{Exp}{[-U(d)/k_{B}T]%
\mathrm{\partial }d}}{\int\limits_{0}^{\infty }\text{Exp}{[-U(d)/k_{B}T]%
\mathrm{\partial }d}}
\end{equation}
The $\bar{d}$ is obtained by averaged over multiple measurements \cite%
{Zha:19}. Here, the offset $\triangle=\bar{d}- d_e$ given in Fig. 5 is
smaller than 1 nm, due to the symmetry of the probability function near the
equilibrium distance. The precise measurement of the thermal Casimir effect relies on the suspended separation of the nanoplate. Fortunately, the high $Q$ spectra in optical cavities offer an opportunity to monitor the separation accurately. As depicted in the insets of Figs. 5(a) and 5(b), the high $Q$-factor is maintained over a wide range of separations.
Compared with previous literature \cite{Rod:10, Est:16}, our work presents an accurate way to detect thermal
Casimir effect via spectroscopic measurements of optical cavities.

\section{Conclusions}

The optical FP cavities tuned by the Casimir forces is
investigated in this work. The system consists of a gold nanoplate
confronted to a Teflon-coated MOS substrate in a liquid environment. The
suspension of the gold nanoplate is dependent on the balance among the
Casimir, gravity and buoyancy forces. One way to modulate the suspension distance of the gold nanoplate is
achieved by electrical gating. The frequency shifting of the reflection spectra
can be tens of nanometers via gating voltages. Furthermore, the control
of the optical resonances via the temperature is also demonstrated. The
temperature modulations can be manifested greatly at sub-micrometer separations. In addition, the
Brownian motion is discussed in different configurations. An new scheme to measure thermal
Casimir effect is suggested by the spectroscopic measurements of the optical FP cavities.

\begin{acknowledgments}
This work is supported by the National Natural Science Foundation of China
(Grant No. 11804288, and No. 61974127), Natural Science Foundation of Henan Province (Grant No. 232300420120), and the Innovation Scientists and Technicians Troop Construction Projects of Henan Province.
\end{acknowledgments}

\bibliography{references}

\begin{thebibliography}{53}
\expandafter\ifx\csname natexlab\endcsname\relax\def\natexlab#1{#1}\fi
\expandafter\ifx\csname bibnamefont\endcsname\relax
  \def\bibnamefont#1{#1}\fi
\expandafter\ifx\csname bibfnamefont\endcsname\relax
  \def\bibfnamefont#1{#1}\fi
\expandafter\ifx\csname citenamefont\endcsname\relax
  \def\citenamefont#1{#1}\fi
\expandafter\ifx\csname url\endcsname\relax
  \def\url#1{\texttt{#1}}\fi
\expandafter\ifx\csname urlprefix\endcsname\relax\def\urlprefix{URL }\fi
\providecommand{\bibinfo}[2]{#2}
\providecommand{\eprint}[2][]{\url{#2}}

\bibitem[{\citenamefont{Casimir}(1948)}]{Cas:48}
\bibinfo{author}{\bibfnamefont{H.~B.~G.} \bibnamefont{Casimir}},
  \bibinfo{journal}{Proc. Kon. Ned. Akad. Wet.} \textbf{\bibinfo{volume}{51}},
  \bibinfo{pages}{793} (\bibinfo{year}{1948}).

\bibitem[{\citenamefont{Lifshitz}(1955)}]{Lif:55}
\bibinfo{author}{\bibfnamefont{E.~M.} \bibnamefont{Lifshitz}},
  \bibinfo{journal}{Zhurnal Eksperimentalnoi Teoreticheskoi Fiziki}
  \textbf{\bibinfo{volume}{29}}, \bibinfo{pages}{94} (\bibinfo{year}{1955}).

\bibitem[{\citenamefont{Woods et~al.}(2016)\citenamefont{Woods, Dalvit,
  Tkatchenko, Rodriguez-Lopez, Rodriguez, and Podgornik}}]{Woo:16}
\bibinfo{author}{\bibfnamefont{L.~M.} \bibnamefont{Woods}},
  \bibinfo{author}{\bibfnamefont{D.~A.~R.} \bibnamefont{Dalvit}},
  \bibinfo{author}{\bibfnamefont{A.}~\bibnamefont{Tkatchenko}},
  \bibinfo{author}{\bibfnamefont{P.}~\bibnamefont{Rodriguez-Lopez}},
  \bibinfo{author}{\bibfnamefont{A.~W.} \bibnamefont{Rodriguez}},
  \bibnamefont{and}
  \bibinfo{author}{\bibfnamefont{R.}~\bibnamefont{Podgornik}},
  \bibinfo{journal}{Rev. Mod. Phys.} \textbf{\bibinfo{volume}{88}},
  \bibinfo{pages}{045003} (\bibinfo{year}{2016}).

\bibitem[{\citenamefont{Kenneth et~al.}(2002)\citenamefont{Kenneth, Klich,
  Mann, and Revzen}}]{Ken:02}
\bibinfo{author}{\bibfnamefont{O.}~\bibnamefont{Kenneth}},
  \bibinfo{author}{\bibfnamefont{I.}~\bibnamefont{Klich}},
  \bibinfo{author}{\bibfnamefont{A.}~\bibnamefont{Mann}}, \bibnamefont{and}
  \bibinfo{author}{\bibfnamefont{M.}~\bibnamefont{Revzen}},
  \bibinfo{journal}{Phys. Rev. Lett.} \textbf{\bibinfo{volume}{89}},
  \bibinfo{pages}{033001} (\bibinfo{year}{2002}).

\bibitem[{\citenamefont{Zhao et~al.}(2009)\citenamefont{Zhao, Zhou, Koschny,
  Economou, and Soukoulis}}]{Zha:09}
\bibinfo{author}{\bibfnamefont{R.}~\bibnamefont{Zhao}},
  \bibinfo{author}{\bibfnamefont{J.}~\bibnamefont{Zhou}},
  \bibinfo{author}{\bibfnamefont{T.}~\bibnamefont{Koschny}},
  \bibinfo{author}{\bibfnamefont{E.}~\bibnamefont{Economou}}, \bibnamefont{and}
  \bibinfo{author}{\bibfnamefont{C.}~\bibnamefont{Soukoulis}},
  \bibinfo{journal}{Phys. Rev. Lett.} \textbf{\bibinfo{volume}{103}},
  \bibinfo{pages}{103602} (\bibinfo{year}{2009}).

\bibitem[{\citenamefont{Levin et~al.}(2010)\citenamefont{Levin, McCauley,
  Rodriguez, Reid, and Johnson}}]{Lev:10}
\bibinfo{author}{\bibfnamefont{M.}~\bibnamefont{Levin}},
  \bibinfo{author}{\bibfnamefont{A.~P.} \bibnamefont{McCauley}},
  \bibinfo{author}{\bibfnamefont{A.~W.} \bibnamefont{Rodriguez}},
  \bibinfo{author}{\bibfnamefont{M.~T.~H.} \bibnamefont{Reid}},
  \bibnamefont{and} \bibinfo{author}{\bibfnamefont{S.~G.}
  \bibnamefont{Johnson}}, \bibinfo{journal}{Phys. Rev. Lett.}
  \textbf{\bibinfo{volume}{105}}, \bibinfo{pages}{090403}
  (\bibinfo{year}{2010}).

\bibitem[{\citenamefont{Rodriguez-Lopez and Grushin}(2014)}]{Rod:14}
\bibinfo{author}{\bibfnamefont{P.}~\bibnamefont{Rodriguez-Lopez}}
  \bibnamefont{and} \bibinfo{author}{\bibfnamefont{A.~G.}
  \bibnamefont{Grushin}}, \bibinfo{journal}{Phys. Rev. Lett.}
  \textbf{\bibinfo{volume}{112}}, \bibinfo{pages}{056804}
  (\bibinfo{year}{2014}).

\bibitem[{\citenamefont{Nie et~al.}(2013)\citenamefont{Nie, Zeng, Lan, and
  Zhu}}]{Nie:13}
\bibinfo{author}{\bibfnamefont{W.}~\bibnamefont{Nie}},
  \bibinfo{author}{\bibfnamefont{R.}~\bibnamefont{Zeng}},
  \bibinfo{author}{\bibfnamefont{Y.}~\bibnamefont{Lan}}, \bibnamefont{and}
  \bibinfo{author}{\bibfnamefont{S.}~\bibnamefont{Zhu}},
  \bibinfo{journal}{Phys. Rev. B} \textbf{\bibinfo{volume}{88}},
  \bibinfo{pages}{085421} (\bibinfo{year}{2013}).

\bibitem[{\citenamefont{Li and Khandekar}(2021)}]{Li:21}
\bibinfo{author}{\bibfnamefont{Z.}~\bibnamefont{Li}} \bibnamefont{and}
  \bibinfo{author}{\bibfnamefont{C.}~\bibnamefont{Khandekar}},
  \bibinfo{journal}{Phys. Rev. Appl.} \textbf{\bibinfo{volume}{16}},
  \bibinfo{pages}{044047} (\bibinfo{year}{2021}).

\bibitem[{\citenamefont{Camacho et~al.}(2022)\citenamefont{Camacho, Gong,
  Spreng, Liberal, Engheta, and Munday}}]{Cam:22}
\bibinfo{author}{\bibfnamefont{M.}~\bibnamefont{Camacho}},
  \bibinfo{author}{\bibfnamefont{T.}~\bibnamefont{Gong}},
  \bibinfo{author}{\bibfnamefont{B.}~\bibnamefont{Spreng}},
  \bibinfo{author}{\bibfnamefont{I.}~\bibnamefont{Liberal}},
  \bibinfo{author}{\bibfnamefont{N.}~\bibnamefont{Engheta}}, \bibnamefont{and}
  \bibinfo{author}{\bibfnamefont{J.~N.} \bibnamefont{Munday}},
  \bibinfo{journal}{Phys. Rev. A} \textbf{\bibinfo{volume}{105}},
  \bibinfo{pages}{L061501} (\bibinfo{year}{2022}).

\bibitem[{\citenamefont{Munday et~al.}(2009)\citenamefont{Munday, Capasso, and
  Parsegian}}]{Mun:09}
\bibinfo{author}{\bibfnamefont{J.~N.} \bibnamefont{Munday}},
  \bibinfo{author}{\bibfnamefont{F.}~\bibnamefont{Capasso}}, \bibnamefont{and}
  \bibinfo{author}{\bibfnamefont{V.~A.} \bibnamefont{Parsegian}},
  \bibinfo{journal}{Nature} \textbf{\bibinfo{volume}{457}},
  \bibinfo{pages}{170} (\bibinfo{year}{2009}).

\bibitem[{\citenamefont{Zhao et~al.}(2019)\citenamefont{Zhao, Li, Yang, Bao,
  Xia, Ashby, Wang, and Zhang}}]{Zha:19}
\bibinfo{author}{\bibfnamefont{R.}~\bibnamefont{Zhao}},
  \bibinfo{author}{\bibfnamefont{L.}~\bibnamefont{Li}},
  \bibinfo{author}{\bibfnamefont{S.}~\bibnamefont{Yang}},
  \bibinfo{author}{\bibfnamefont{W.}~\bibnamefont{Bao}},
  \bibinfo{author}{\bibfnamefont{Y.}~\bibnamefont{Xia}},
  \bibinfo{author}{\bibfnamefont{P.}~\bibnamefont{Ashby}},
  \bibinfo{author}{\bibfnamefont{Y.}~\bibnamefont{Wang}}, \bibnamefont{and}
  \bibinfo{author}{\bibfnamefont{X.}~\bibnamefont{Zhang}},
  \bibinfo{journal}{Science} \textbf{\bibinfo{volume}{364}},
  \bibinfo{pages}{984} (\bibinfo{year}{2019}).

\bibitem[{\citenamefont{Dou et~al.}(2014)\citenamefont{Dou, Lou, Bostr{\"o}m,
  Brevik, and Persson}}]{Dou:14}
\bibinfo{author}{\bibfnamefont{M.}~\bibnamefont{Dou}},
  \bibinfo{author}{\bibfnamefont{F.}~\bibnamefont{Lou}},
  \bibinfo{author}{\bibfnamefont{M.}~\bibnamefont{Bostr{\"o}m}},
  \bibinfo{author}{\bibfnamefont{I.}~\bibnamefont{Brevik}}, \bibnamefont{and}
  \bibinfo{author}{\bibfnamefont{C.}~\bibnamefont{Persson}},
  \bibinfo{journal}{Phys. Rev. B} \textbf{\bibinfo{volume}{89}},
  \bibinfo{pages}{201407(R)} (\bibinfo{year}{2014}).

\bibitem[{\citenamefont{Esteso et~al.}(2015)\citenamefont{Esteso,
  Carretero-Palacios, and M{\'\i}guez}}]{Est:15}
\bibinfo{author}{\bibfnamefont{V.}~\bibnamefont{Esteso}},
  \bibinfo{author}{\bibfnamefont{S.}~\bibnamefont{Carretero-Palacios}},
  \bibnamefont{and}
  \bibinfo{author}{\bibfnamefont{H.}~\bibnamefont{M{\'\i}guez}},
  \bibinfo{journal}{J. Phys. Chem. C} \textbf{\bibinfo{volume}{119}},
  \bibinfo{pages}{5663} (\bibinfo{year}{2015}).

\bibitem[{\citenamefont{Liu and Zhang}(2016)}]{Liu:16}
\bibinfo{author}{\bibfnamefont{X.}~\bibnamefont{Liu}} \bibnamefont{and}
  \bibinfo{author}{\bibfnamefont{Z.~M.} \bibnamefont{Zhang}},
  \bibinfo{journal}{Phys. Rev. Appl.} \textbf{\bibinfo{volume}{5}},
  \bibinfo{pages}{034004} (\bibinfo{year}{2016}).

\bibitem[{\citenamefont{Ye et~al.}(2018)\citenamefont{Ye, Hu, Zhao, and
  Meng}}]{Ye:18}
\bibinfo{author}{\bibfnamefont{Y.}~\bibnamefont{Ye}},
  \bibinfo{author}{\bibfnamefont{Q.}~\bibnamefont{Hu}},
  \bibinfo{author}{\bibfnamefont{Q.}~\bibnamefont{Zhao}}, \bibnamefont{and}
  \bibinfo{author}{\bibfnamefont{Y.}~\bibnamefont{Meng}},
  \bibinfo{journal}{Phys. Rev. B} \textbf{\bibinfo{volume}{98}},
  \bibinfo{pages}{035410} (\bibinfo{year}{2018}).

\bibitem[{\citenamefont{Esteso et~al.}(2022)\citenamefont{Esteso,
  Carretero-Palacios, and M{\'\i}guez}}]{Est:22}
\bibinfo{author}{\bibfnamefont{V.}~\bibnamefont{Esteso}},
  \bibinfo{author}{\bibfnamefont{S.}~\bibnamefont{Carretero-Palacios}},
  \bibnamefont{and}
  \bibinfo{author}{\bibfnamefont{H.}~\bibnamefont{M{\'\i}guez}},
  \bibinfo{journal}{J. Phys. Chem. Lett.} \textbf{\bibinfo{volume}{13}},
  \bibinfo{pages}{4513} (\bibinfo{year}{2022}).

\bibitem[{\citenamefont{Esteso et~al.}(2019)\citenamefont{Esteso,
  Carretero-Palacios, and M{\'\i}guez}}]{Est:19}
\bibinfo{author}{\bibfnamefont{V.}~\bibnamefont{Esteso}},
  \bibinfo{author}{\bibfnamefont{S.}~\bibnamefont{Carretero-Palacios}},
  \bibnamefont{and}
  \bibinfo{author}{\bibfnamefont{H.}~\bibnamefont{M{\'\i}guez}},
  \bibinfo{journal}{J. Phys. Chem. Lett.} \textbf{\bibinfo{volume}{10}},
  \bibinfo{pages}{5856} (\bibinfo{year}{2019}).

\bibitem[{\citenamefont{Vaughan}(2017)}]{Vau:17}
\bibinfo{author}{\bibfnamefont{J.~M.} \bibnamefont{Vaughan}},
  \emph{\bibinfo{title}{The Fabry--Perot interferometer: history, theory,
  practice and applications}} (\bibinfo{publisher}{Routledge},
  \bibinfo{year}{2017}).

\bibitem[{\citenamefont{Caligiuri et~al.}(2020)\citenamefont{Caligiuri, Biffi,
  Palei, Mart{\'\i}n-Garc{\'\i}a, Pothuraju, Bretonni{\`e}re, and
  Krahne}}]{Cal:20}
\bibinfo{author}{\bibfnamefont{V.}~\bibnamefont{Caligiuri}},
  \bibinfo{author}{\bibfnamefont{G.}~\bibnamefont{Biffi}},
  \bibinfo{author}{\bibfnamefont{M.}~\bibnamefont{Palei}},
  \bibinfo{author}{\bibfnamefont{B.}~\bibnamefont{Mart{\'\i}n-Garc{\'\i}a}},
  \bibinfo{author}{\bibfnamefont{R.~D.} \bibnamefont{Pothuraju}},
  \bibinfo{author}{\bibfnamefont{Y.}~\bibnamefont{Bretonni{\`e}re}},
  \bibnamefont{and} \bibinfo{author}{\bibfnamefont{R.}~\bibnamefont{Krahne}},
  \bibinfo{journal}{Adv. Opt. Mater.} \textbf{\bibinfo{volume}{8}},
  \bibinfo{pages}{1901215} (\bibinfo{year}{2020}).

\bibitem[{\citenamefont{Liu et~al.}(2010)\citenamefont{Liu, Mesch, Weiss,
  Hentschel, and Giessen}}]{Liu:10}
\bibinfo{author}{\bibfnamefont{N.}~\bibnamefont{Liu}},
  \bibinfo{author}{\bibfnamefont{M.}~\bibnamefont{Mesch}},
  \bibinfo{author}{\bibfnamefont{T.}~\bibnamefont{Weiss}},
  \bibinfo{author}{\bibfnamefont{M.}~\bibnamefont{Hentschel}},
  \bibnamefont{and} \bibinfo{author}{\bibfnamefont{H.}~\bibnamefont{Giessen}},
  \bibinfo{journal}{Nano Lett.} \textbf{\bibinfo{volume}{10}},
  \bibinfo{pages}{2342} (\bibinfo{year}{2010}).

\bibitem[{\citenamefont{Deng et~al.}(2015)\citenamefont{Deng, Li, Stan,
  Rosenmann, Czaplewski, Gao, and Yang}}]{Den:15}
\bibinfo{author}{\bibfnamefont{H.}~\bibnamefont{Deng}},
  \bibinfo{author}{\bibfnamefont{Z.}~\bibnamefont{Li}},
  \bibinfo{author}{\bibfnamefont{L.}~\bibnamefont{Stan}},
  \bibinfo{author}{\bibfnamefont{D.}~\bibnamefont{Rosenmann}},
  \bibinfo{author}{\bibfnamefont{D.}~\bibnamefont{Czaplewski}},
  \bibinfo{author}{\bibfnamefont{J.}~\bibnamefont{Gao}}, \bibnamefont{and}
  \bibinfo{author}{\bibfnamefont{X.}~\bibnamefont{Yang}},
  \bibinfo{journal}{Opt. Lett.} \textbf{\bibinfo{volume}{40}},
  \bibinfo{pages}{2592} (\bibinfo{year}{2015}).

\bibitem[{\citenamefont{Ge et~al.}(2020{\natexlab{a}})\citenamefont{Ge, Shi,
  Liu, and Gong}}]{Ge:20b}
\bibinfo{author}{\bibfnamefont{L.}~\bibnamefont{Ge}},
  \bibinfo{author}{\bibfnamefont{X.}~\bibnamefont{Shi}},
  \bibinfo{author}{\bibfnamefont{L.}~\bibnamefont{Liu}}, \bibnamefont{and}
  \bibinfo{author}{\bibfnamefont{K.}~\bibnamefont{Gong}},
  \bibinfo{journal}{Phys. Rev. B} \textbf{\bibinfo{volume}{102}},
  \bibinfo{pages}{075428} (\bibinfo{year}{2020}{\natexlab{a}}).

\bibitem[{\citenamefont{Gong et~al.}(2022)\citenamefont{Gong, Spreng, Camacho,
  Liberal, Engheta, and Munday}}]{Gon:22}
\bibinfo{author}{\bibfnamefont{T.}~\bibnamefont{Gong}},
  \bibinfo{author}{\bibfnamefont{B.}~\bibnamefont{Spreng}},
  \bibinfo{author}{\bibfnamefont{M.}~\bibnamefont{Camacho}},
  \bibinfo{author}{\bibfnamefont{I.}~\bibnamefont{Liberal}},
  \bibinfo{author}{\bibfnamefont{N.}~\bibnamefont{Engheta}}, \bibnamefont{and}
  \bibinfo{author}{\bibfnamefont{J.~N.} \bibnamefont{Munday}},
  \bibinfo{journal}{Phys. Rev. A} \textbf{\bibinfo{volume}{106}},
  \bibinfo{pages}{062824} (\bibinfo{year}{2022}).

\bibitem[{\citenamefont{Jiang and Wilczek}(2019)}]{Jia:19}
\bibinfo{author}{\bibfnamefont{Q.-D.} \bibnamefont{Jiang}} \bibnamefont{and}
  \bibinfo{author}{\bibfnamefont{F.}~\bibnamefont{Wilczek}},
  \bibinfo{journal}{Phys. Rev. B} \textbf{\bibinfo{volume}{99}},
  \bibinfo{pages}{125403} (\bibinfo{year}{2019}).

\bibitem[{\citenamefont{Zeng and Yang}(2011)}]{Zen:11}
\bibinfo{author}{\bibfnamefont{R.}~\bibnamefont{Zeng}} \bibnamefont{and}
  \bibinfo{author}{\bibfnamefont{Y.}~\bibnamefont{Yang}},
  \bibinfo{journal}{Phys. Rev. A} \textbf{\bibinfo{volume}{83}},
  \bibinfo{pages}{012517} (\bibinfo{year}{2011}).

\bibitem[{\citenamefont{Wang et~al.}(2006)\citenamefont{Wang, Zhang, Pei, and
  Liu}}]{Wan:06}
\bibinfo{author}{\bibfnamefont{J.}~\bibnamefont{Wang}},
  \bibinfo{author}{\bibfnamefont{X.}~\bibnamefont{Zhang}},
  \bibinfo{author}{\bibfnamefont{S.-Y.} \bibnamefont{Pei}}, \bibnamefont{and}
  \bibinfo{author}{\bibfnamefont{D.-H.} \bibnamefont{Liu}},
  \bibinfo{journal}{Phys. Rev. A} \textbf{\bibinfo{volume}{73}},
  \bibinfo{pages}{042103} (\bibinfo{year}{2006}).

\bibitem[{\citenamefont{Chen et~al.}(2007)\citenamefont{Chen, Klimchitskaya,
  Mostepanenko, and Mohideen}}]{Che:07}
\bibinfo{author}{\bibfnamefont{F.}~\bibnamefont{Chen}},
  \bibinfo{author}{\bibfnamefont{G.~L.} \bibnamefont{Klimchitskaya}},
  \bibinfo{author}{\bibfnamefont{V.~M.} \bibnamefont{Mostepanenko}},
  \bibnamefont{and} \bibinfo{author}{\bibfnamefont{U.}~\bibnamefont{Mohideen}},
  \bibinfo{journal}{Phys. Rev. B} \textbf{\bibinfo{volume}{76}},
  \bibinfo{pages}{035338} (\bibinfo{year}{2007}).

\bibitem[{\citenamefont{Chang et~al.}(2011)\citenamefont{Chang, Banishev,
  Klimchitskaya, Mostepanenko, and Mohideen}}]{Cha:11}
\bibinfo{author}{\bibfnamefont{C.-C.} \bibnamefont{Chang}},
  \bibinfo{author}{\bibfnamefont{A.~A.} \bibnamefont{Banishev}},
  \bibinfo{author}{\bibfnamefont{G.~L.} \bibnamefont{Klimchitskaya}},
  \bibinfo{author}{\bibfnamefont{V.~M.} \bibnamefont{Mostepanenko}},
  \bibnamefont{and} \bibinfo{author}{\bibfnamefont{U.}~\bibnamefont{Mohideen}},
  \bibinfo{journal}{Phys. Rev. Lett.} \textbf{\bibinfo{volume}{107}},
  \bibinfo{pages}{090403} (\bibinfo{year}{2011}).

\bibitem[{\citenamefont{Yampol’skii et~al.}(2008)\citenamefont{Yampol’skii,
  Savel’ev, Mayselis, Apostolov, and Nori}}]{Yam:08}
\bibinfo{author}{\bibfnamefont{V.~A.} \bibnamefont{Yampol’skii}},
  \bibinfo{author}{\bibfnamefont{S.}~\bibnamefont{Savel’ev}},
  \bibinfo{author}{\bibfnamefont{Z.~A.} \bibnamefont{Mayselis}},
  \bibinfo{author}{\bibfnamefont{S.~S.} \bibnamefont{Apostolov}},
  \bibnamefont{and} \bibinfo{author}{\bibfnamefont{F.}~\bibnamefont{Nori}},
  \bibinfo{journal}{Phys. Rev.Lett.} \textbf{\bibinfo{volume}{101}},
  \bibinfo{pages}{096803} (\bibinfo{year}{2008}).

\bibitem[{\citenamefont{Galkina et~al.}(2009)\citenamefont{Galkina, Ivanov,
  Savel'ev, Yampol'skii, and Nori}}]{Gal:09}
\bibinfo{author}{\bibfnamefont{E.~G.} \bibnamefont{Galkina}},
  \bibinfo{author}{\bibfnamefont{B.~A.} \bibnamefont{Ivanov}},
  \bibinfo{author}{\bibfnamefont{S.}~\bibnamefont{Savel'ev}},
  \bibinfo{author}{\bibfnamefont{V.~A.} \bibnamefont{Yampol'skii}},
  \bibnamefont{and} \bibinfo{author}{\bibfnamefont{F.}~\bibnamefont{Nori}},
  \bibinfo{journal}{Phys. Rev. B} \textbf{\bibinfo{volume}{80}},
  \bibinfo{pages}{125119} (\bibinfo{year}{2009}).

\bibitem[{\citenamefont{Bostr{\"o}m et~al.}(2018)\citenamefont{Bostr{\"o}m,
  Dou, Malyi, Parashar, Parsons, Brevik, and Persson}}]{Bos:18}
\bibinfo{author}{\bibfnamefont{M.}~\bibnamefont{Bostr{\"o}m}},
  \bibinfo{author}{\bibfnamefont{M.}~\bibnamefont{Dou}},
  \bibinfo{author}{\bibfnamefont{O.~I.} \bibnamefont{Malyi}},
  \bibinfo{author}{\bibfnamefont{P.}~\bibnamefont{Parashar}},
  \bibinfo{author}{\bibfnamefont{D.~F.} \bibnamefont{Parsons}},
  \bibinfo{author}{\bibfnamefont{I.}~\bibnamefont{Brevik}}, \bibnamefont{and}
  \bibinfo{author}{\bibfnamefont{C.}~\bibnamefont{Persson}},
  \bibinfo{journal}{Phys. Rev. B} \textbf{\bibinfo{volume}{97}},
  \bibinfo{pages}{125421} (\bibinfo{year}{2018}).

\bibitem[{\citenamefont{Ge and Shi}(2022)}]{Ge:22}
\bibinfo{author}{\bibfnamefont{L.}~\bibnamefont{Ge}} \bibnamefont{and}
  \bibinfo{author}{\bibfnamefont{X.}~\bibnamefont{Shi}},
  \bibinfo{journal}{Phys. Lett. A} \textbf{\bibinfo{volume}{450}},
  \bibinfo{pages}{128392} (\bibinfo{year}{2022}).

\bibitem[{\citenamefont{Bostr{\"o}m and Sernelius}(2000)}]{Mat:00}
\bibinfo{author}{\bibfnamefont{M.}~\bibnamefont{Bostr{\"o}m}} \bibnamefont{and}
  \bibinfo{author}{\bibfnamefont{B.~E.} \bibnamefont{Sernelius}},
  \bibinfo{journal}{Phys. Rev. Lett.} \textbf{\bibinfo{volume}{84}},
  \bibinfo{pages}{4757} (\bibinfo{year}{2000}).

\bibitem[{\citenamefont{Sushkov et~al.}(2011)\citenamefont{Sushkov, Kim,
  Dalvit, and Lamoreaux}}]{Sus:11}
\bibinfo{author}{\bibfnamefont{A.}~\bibnamefont{Sushkov}},
  \bibinfo{author}{\bibfnamefont{W.}~\bibnamefont{Kim}},
  \bibinfo{author}{\bibfnamefont{D.}~\bibnamefont{Dalvit}}, \bibnamefont{and}
  \bibinfo{author}{\bibfnamefont{S.}~\bibnamefont{Lamoreaux}},
  \bibinfo{journal}{Nat. Phys.} \textbf{\bibinfo{volume}{7}},
  \bibinfo{pages}{230} (\bibinfo{year}{2011}).

\bibitem[{\citenamefont{Liu et~al.}(2021)\citenamefont{Liu, Zhang,
  Klimchitskaya, Mostepanenko, and Mohideen}}]{Liu:21}
\bibinfo{author}{\bibfnamefont{M.}~\bibnamefont{Liu}},
  \bibinfo{author}{\bibfnamefont{Y.}~\bibnamefont{Zhang}},
  \bibinfo{author}{\bibfnamefont{G.~L.} \bibnamefont{Klimchitskaya}},
  \bibinfo{author}{\bibfnamefont{V.~M.} \bibnamefont{Mostepanenko}},
  \bibnamefont{and} \bibinfo{author}{\bibfnamefont{U.}~\bibnamefont{Mohideen}},
  \bibinfo{journal}{Phys. Rev. B} \textbf{\bibinfo{volume}{104}},
  \bibinfo{pages}{085436} (\bibinfo{year}{2021}).

\bibitem[{\citenamefont{Abbas et~al.}(2017)\citenamefont{Abbas, Guizal, and
  Antezza}}]{Abb:17}
\bibinfo{author}{\bibfnamefont{C.}~\bibnamefont{Abbas}},
  \bibinfo{author}{\bibfnamefont{B.}~\bibnamefont{Guizal}}, \bibnamefont{and}
  \bibinfo{author}{\bibfnamefont{M.}~\bibnamefont{Antezza}},
  \bibinfo{journal}{Phys. Rev. Lett.} \textbf{\bibinfo{volume}{118}},
  \bibinfo{pages}{126101} (\bibinfo{year}{2017}).

\bibitem[{\citenamefont{Bimonte et~al.}(2017)\citenamefont{Bimonte,
  Klimchitskaya, and Mostepanenko}}]{Bim:17}
\bibinfo{author}{\bibfnamefont{G.}~\bibnamefont{Bimonte}},
  \bibinfo{author}{\bibfnamefont{G.~L.} \bibnamefont{Klimchitskaya}},
  \bibnamefont{and} \bibinfo{author}{\bibfnamefont{V.~M.}
  \bibnamefont{Mostepanenko}}, \bibinfo{journal}{Phys. Rev. A}
  \textbf{\bibinfo{volume}{96}}, \bibinfo{pages}{012517}
  (\bibinfo{year}{2017}).

\bibitem[{\citenamefont{Khusnutdinov et~al.}(2018)\citenamefont{Khusnutdinov,
  Kashapov, and Woods}}]{Khu:18}
\bibinfo{author}{\bibfnamefont{N.}~\bibnamefont{Khusnutdinov}},
  \bibinfo{author}{\bibfnamefont{R.}~\bibnamefont{Kashapov}}, \bibnamefont{and}
  \bibinfo{author}{\bibfnamefont{L.~M.} \bibnamefont{Woods}},
  \bibinfo{journal}{2D Mater.} \textbf{\bibinfo{volume}{5}},
  \bibinfo{pages}{035032} (\bibinfo{year}{2018}).

\bibitem[{\citenamefont{Klimchitskaya and Mostepanenko}(2015)}]{Kli:15}
\bibinfo{author}{\bibfnamefont{G.~L.} \bibnamefont{Klimchitskaya}}
  \bibnamefont{and} \bibinfo{author}{\bibfnamefont{V.~M.}
  \bibnamefont{Mostepanenko}}, \bibinfo{journal}{Phys. Rev. B}
  \textbf{\bibinfo{volume}{91}}, \bibinfo{pages}{174501}
  (\bibinfo{year}{2015}).

\bibitem[{\citenamefont{Ge et~al.}(2020{\natexlab{b}})\citenamefont{Ge, Shi,
  Xu, and Gong}}]{Ge:20a}
\bibinfo{author}{\bibfnamefont{L.}~\bibnamefont{Ge}},
  \bibinfo{author}{\bibfnamefont{X.}~\bibnamefont{Shi}},
  \bibinfo{author}{\bibfnamefont{Z.}~\bibnamefont{Xu}}, \bibnamefont{and}
  \bibinfo{author}{\bibfnamefont{K.}~\bibnamefont{Gong}},
  \bibinfo{journal}{Phys. Rev. B} \textbf{\bibinfo{volume}{101}},
  \bibinfo{pages}{104107} (\bibinfo{year}{2020}{\natexlab{b}}).

\bibitem[{\citenamefont{Moazzami~Gudarzi and Aboutalebi}(2021)}]{Moa:21}
\bibinfo{author}{\bibfnamefont{M.}~\bibnamefont{Moazzami~Gudarzi}}
  \bibnamefont{and} \bibinfo{author}{\bibfnamefont{S.~H.}
  \bibnamefont{Aboutalebi}}, \bibinfo{journal}{Sci. Adv.}
  \textbf{\bibinfo{volume}{7}}, \bibinfo{pages}{eabg2272}
  (\bibinfo{year}{2021}).

\bibitem[{\citenamefont{Sehmi et~al.}(2017)\citenamefont{Sehmi, Langbein, and
  Muljarov}}]{Seh:17}
\bibinfo{author}{\bibfnamefont{H.~S.} \bibnamefont{Sehmi}},
  \bibinfo{author}{\bibfnamefont{W.}~\bibnamefont{Langbein}}, \bibnamefont{and}
  \bibinfo{author}{\bibfnamefont{E.~A.} \bibnamefont{Muljarov}},
  \bibinfo{journal}{Phys. Rev. B} \textbf{\bibinfo{volume}{95}},
  \bibinfo{pages}{115444} (\bibinfo{year}{2017}).

\bibitem[{\citenamefont{Krasavin and Zayats}(2012)}]{Kra:12}
\bibinfo{author}{\bibfnamefont{A.~V.} \bibnamefont{Krasavin}} \bibnamefont{and}
  \bibinfo{author}{\bibfnamefont{A.~V.} \bibnamefont{Zayats}},
  \bibinfo{journal}{Phys. Rev. Lett.} \textbf{\bibinfo{volume}{109}},
  \bibinfo{pages}{053901} (\bibinfo{year}{2012}).

\bibitem[{\citenamefont{Jellison~Jr and Modine}(1996)}]{Jel:96}
\bibinfo{author}{\bibfnamefont{G.}~\bibnamefont{Jellison~Jr}} \bibnamefont{and}
  \bibinfo{author}{\bibfnamefont{F.}~\bibnamefont{Modine}},
  \bibinfo{journal}{Appl. Phys. Lett.} \textbf{\bibinfo{volume}{69}},
  \bibinfo{pages}{371} (\bibinfo{year}{1996}).

\bibitem[{\citenamefont{Banishev et~al.}(2012)\citenamefont{Banishev, Chang,
  Castillo-Garza, Klimchitskaya, Mostepanenko, and Mohideen}}]{Ban:12}
\bibinfo{author}{\bibfnamefont{A.~A.} \bibnamefont{Banishev}},
  \bibinfo{author}{\bibfnamefont{C.-C.} \bibnamefont{Chang}},
  \bibinfo{author}{\bibfnamefont{R.}~\bibnamefont{Castillo-Garza}},
  \bibinfo{author}{\bibfnamefont{G.~L.} \bibnamefont{Klimchitskaya}},
  \bibinfo{author}{\bibfnamefont{V.~M.} \bibnamefont{Mostepanenko}},
  \bibnamefont{and} \bibinfo{author}{\bibfnamefont{U.}~\bibnamefont{Mohideen}},
  \bibinfo{journal}{Phys. Rev. B} \textbf{\bibinfo{volume}{85}},
  \bibinfo{pages}{045436} (\bibinfo{year}{2012}).

\bibitem[{\citenamefont{Yi et~al.}(2013)\citenamefont{Yi, Shim, Zhu, Zhu, Reed,
  and Cubukcu}}]{Yi:13}
\bibinfo{author}{\bibfnamefont{F.}~\bibnamefont{Yi}},
  \bibinfo{author}{\bibfnamefont{E.}~\bibnamefont{Shim}},
  \bibinfo{author}{\bibfnamefont{A.~Y.} \bibnamefont{Zhu}},
  \bibinfo{author}{\bibfnamefont{H.}~\bibnamefont{Zhu}},
  \bibinfo{author}{\bibfnamefont{J.~C.} \bibnamefont{Reed}}, \bibnamefont{and}
  \bibinfo{author}{\bibfnamefont{E.}~\bibnamefont{Cubukcu}},
  \bibinfo{journal}{Appl. Phys. Lett.} \textbf{\bibinfo{volume}{102}},
  \bibinfo{pages}{221102} (\bibinfo{year}{2013}).

\bibitem[{\citenamefont{Papadakis and Atwater}(2015)}]{Pap:15}
\bibinfo{author}{\bibfnamefont{G.~T.} \bibnamefont{Papadakis}}
  \bibnamefont{and} \bibinfo{author}{\bibfnamefont{H.~A.}
  \bibnamefont{Atwater}}, \bibinfo{journal}{Phys. Rev. B}
  \textbf{\bibinfo{volume}{92}}, \bibinfo{pages}{184101}
  (\bibinfo{year}{2015}).

\bibitem[{\citenamefont{Sire et~al.}(2007)\citenamefont{Sire, Blonkowski,
  Gordon, and Baron}}]{Sir:07}
\bibinfo{author}{\bibfnamefont{C.}~\bibnamefont{Sire}},
  \bibinfo{author}{\bibfnamefont{S.}~\bibnamefont{Blonkowski}},
  \bibinfo{author}{\bibfnamefont{M.~J.} \bibnamefont{Gordon}},
  \bibnamefont{and} \bibinfo{author}{\bibfnamefont{T.}~\bibnamefont{Baron}},
  \bibinfo{journal}{Appl. Phys. Lett.} \textbf{\bibinfo{volume}{91}},
  \bibinfo{pages}{242905} (\bibinfo{year}{2007}).

\bibitem[{\citenamefont{Esteso et~al.}(2016)\citenamefont{Esteso,
  Carretero-Palacios, and M{\'\i}guez}}]{Est:16}
\bibinfo{author}{\bibfnamefont{V.}~\bibnamefont{Esteso}},
  \bibinfo{author}{\bibfnamefont{S.}~\bibnamefont{Carretero-Palacios}},
  \bibnamefont{and}
  \bibinfo{author}{\bibfnamefont{H.}~\bibnamefont{M{\'\i}guez}},
  \bibinfo{journal}{J. Appl. Phys.} \textbf{\bibinfo{volume}{119}},
  \bibinfo{pages}{144301} (\bibinfo{year}{2016}).

\bibitem[{\citenamefont{Rodriguez et~al.}(2010)\citenamefont{Rodriguez, Woolf,
  McCauley, Capasso, Joannopoulos, and Johnson}}]{Rod:10}
\bibinfo{author}{\bibfnamefont{A.~W.} \bibnamefont{Rodriguez}},
  \bibinfo{author}{\bibfnamefont{D.}~\bibnamefont{Woolf}},
  \bibinfo{author}{\bibfnamefont{A.~P.} \bibnamefont{McCauley}},
  \bibinfo{author}{\bibfnamefont{F.}~\bibnamefont{Capasso}},
  \bibinfo{author}{\bibfnamefont{J.~D.} \bibnamefont{Joannopoulos}},
  \bibnamefont{and} \bibinfo{author}{\bibfnamefont{S.~G.}
  \bibnamefont{Johnson}}, \bibinfo{journal}{Phys. Rev. Lett.}
  \textbf{\bibinfo{volume}{105}}, \bibinfo{pages}{060401}
  (\bibinfo{year}{2010}).

\bibitem[{\citenamefont{Zhan et~al.}(2013)\citenamefont{Zhan, Shi, Dai, Liu,
  and Zi}}]{Zha:13}
\bibinfo{author}{\bibfnamefont{T.}~\bibnamefont{Zhan}},
  \bibinfo{author}{\bibfnamefont{X.}~\bibnamefont{Shi}},
  \bibinfo{author}{\bibfnamefont{Y.}~\bibnamefont{Dai}},
  \bibinfo{author}{\bibfnamefont{X.}~\bibnamefont{Liu}}, \bibnamefont{and}
  \bibinfo{author}{\bibfnamefont{J.}~\bibnamefont{Zi}}, \bibinfo{journal}{J.
  Phys.: Condens. Matter} \textbf{\bibinfo{volume}{25}},
  \bibinfo{pages}{215301} (\bibinfo{year}{2013}).

\bibitem[{\citenamefont{Varela et~al.}(2011)\citenamefont{Varela, Rodriguez,
  McCauley, and Johnson}}]{Var:11}
\bibinfo{author}{\bibfnamefont{J.}~\bibnamefont{Varela}},
  \bibinfo{author}{\bibfnamefont{A.~W.} \bibnamefont{Rodriguez}},
  \bibinfo{author}{\bibfnamefont{A.~P.} \bibnamefont{McCauley}},
  \bibnamefont{and} \bibinfo{author}{\bibfnamefont{S.~G.}
  \bibnamefont{Johnson}}, \bibinfo{journal}{Phys. Rev. A}
  \textbf{\bibinfo{volume}{83}}, \bibinfo{pages}{042516}
  (\bibinfo{year}{2011}).

\end{thebibliography}

\end{document}